\documentclass[namedreferences]{kluwer}
\usepackage[dvips]{graphicx}
\begin{document}
\begin{article}

\begin{opening}
\title{Evidence for Flash Mixing in He-rich sdB Stars}
\author{Allen V. \surname{Sweigart}}
\author{Thierry \surname{Lanz}}
\institute{NASA Goddard Space Flight Center, Code 681, Greenbelt, MD 20771, USA}
\author{Thomas M. \surname{Brown}}
\institute{Space Telescope Science Institute, 3700 San Martin Drive, Baltimore, MD 21218, USA}
\author{Ivan \surname{Hubeny}}
\institute{National Optical Astronomy Observatory, PO Box 26732, Tucson, AZ 85726, USA}
\author{Wayne B. \surname{Landsman}}
\institute{NASA Goddard Space Flight Center, Code 681, Greenbelt, MD 20771, USA}
\runningauthor{Sweigart, Lanz, Brown, Hubeny \& Landsman}
\runningtitle{Flash Mixing in He-rich sdB Stars}

\begin{abstract}
We present FUSE spectra of three He-rich sdB stars.  Two of these
stars, PG1544+488 and JL87, reveal extremely strong C III lines,
suggesting that they have mixed triple-$\alpha$ carbon from
the deep interior out to their surfaces.  Using TLUSTY NLTE
line-blanketed model atmospheres, we find that PG1544+488
has a surface composition of 96\% He, 2\% C,
and 1\% N.  JL87 shows a similar surface enrichment of C
and N but still retains a significant amount of
hydrogen.  In contrast, the third star, LB1766, is
devoid of hydrogen and strongly depleted of carbon, indicating
that its surface material has undergone CN-cycle processing.

We interpret these observations with new evolutionary
calculations which suggest that He-rich sdB stars with C-rich
compositions arise from a delayed helium-core flash on the
white-dwarf cooling curve.  During such a flash the interior
convection zone will penetrate into the stellar envelope,
thereby mixing the envelope with the He- and C-rich core.  Such
``flash-mixed'' stars will arrive on the extreme horizontal branch
(EHB) with He- and C-rich surface compositions and will be hotter
than the hottest canonical EHB stars.  Two types of flash mixing
are possible: ``deep'' and ``shallow'', depending on whether the
hydrogen envelope is mixed deeply into the site of the helium
flash or only with the outer layers of the core.  Based
on both their stellar parameters
and surface compositions, we suggest that PG1544+488 and JL87 are
examples of ``deep'' and ``shallow'' flash mixing, respectively.
\end{abstract}

\end{opening}

\section{Introduction}

Extreme horizontal-branch (EHB) stars occupy the hot end of the
horizontal branch (HB) and are distinguished by their high effective
temperatures, $\rm T_{eff} > 20,000$~K, and high surface
gravities, log g $>$ 5.  In the Galactic field they correspond to
the subdwarf B (sdB) stars, while in the globular clusters they lie
at the faint end of the blue HB tail.  Most sdB stars are extremely
deficient in helium, but a minority are
helium-rich.  Deficiencies in helium have been
attributed to gravitational settling, but the helium-rich members
of the class, especially those with enhanced carbon,
present a puzzling exception.

Here we explore the origins of the He-rich sdB stars in light of
new Far Ultraviolet Spectroscopic Explorer (FUSE) observations
of three field He-sdB stars and new theoretical calculations for
the helium flash.

\section{FUSE Observations and Spectroscopic Analysis}

We selected three He-rich sdB stars previously studied with optical
spectroscopy: PG1544+488, JL87 and LB1766.  PG1544+488 is
the class prototype and is nearly devoid of hydrogen.  JL87
is moderately enriched in helium, He/H $ \sim $ 0.1 - 0.2
by number.  Optical spectra suggest that these two stars have
similar effective temperatures $\rm T_{eff} \sim 30,000$~K.  LB1766
shows no evidence for hydrogen.  Unlike the other two stars,
its spectrum does not suggest strong carbon enrichment.

We analyzed our FUSE spectra with NLTE line-blanketed TLUSTY model
atmospheres which included 741 individual levels of H~I-II,
He~I-III, C~II-V, N~II-VI and Si~III-V grouped into 301 NLTE
superlevels.  We used lines of minor ions (Si~III, C~II, C~IV) to
determine $\rm T_{eff}$ and lines of dominant ions (C~III, N~III, Si~IV)
to derive the surface abundances.  Surface gravities were derived
by matching the wings of the Lyman and He~II lines.  Interstellar
line absorption was added to the photospheric spectrum in order
to match the interstellar features.

\begin{table}[b]
\caption[]{Results from analysis of FUSE spectra}
\begin{tabular}{llll}
\hline
\multicolumn{1}{c}{Star} & \multicolumn{1}{c}{PG1544$+$488} & \multicolumn{1}{c}{JL87} & \multicolumn{1}{c}{LB1766} \\
\hline
Stellar parameters: \\
$\rm~~~~T_{eff}$ [K] & 36000$\; \pm \;$2000 & 29000$\; \pm \;$2000 & 40000$\; \pm \;$2000 \\
$~~~ \;$log g [cm/$\rm s^2$] & 6.0$\; \pm \;$0.3 & 5.5$\; \pm \;$0.3 & 6.3$\; \pm \;$0.3 \\
\multicolumn{2}{l}{Surface abundances (mass fraction):} \\
$~~~~$H & $< \:$0.002 & 0.55 $-$ 0.70 & 0.0025 \\
$~~~~$He & 0.96 & 0.43 $-$ 0.28 & 0.99 \\
$~~~~$C & 0.02 & 0.014 & 0.0001 \\
$~~~~$N & 0.01 & 0.004 & 0.006 \\
\hline
\end{tabular}
\end{table}

Our results, given in Table 1, confirm the high C abundances of
PG1544+488 and JL87 suggested by previous optical studies.  Both stars
show evidence for the dredge-up of triple-$\alpha$ C.  The surface
of PG1544+488 is virtually devoid of H, while JL87 has a significant
amount of H in its atmosphere.  The third star LB1766 is also devoid
of H like PG1544+488, but its atmosphere is strongly depleted
in C.  The surface composition of LB1766 is consistent with
CN-cycle processing.

\section{Evolutionary Interpretation: Flash Mixing}

Low-mass stars which undergo very high mass loss on the red-giant
branch can evolve to high effective temperatures before igniting
helium in their cores, leading to the so-called ``hot flashers'' (see,
e.g., \opencite{Brown01}).  The high helium-burning
luminosity during the helium flash ($\gtrsim 10^9~L_\odot$)
produces a temporary convection zone that extends from the site of the
helium flash outward to just inside the base of the hydrogen envelope
(see Figure 1).  Following the flash peak, this convection zone retreats
and disappears although a small convective shell persists in the outer
part of the core for a few thousand years.  Normally the flash convection
zone does not penetrate into the envelope due to the high entropy
barrier of the hydrogen shell.  However, such penetration becomes
inevitable if the helium flash occurs on the white-dwarf cooling
curve, where the hydrogen shell is much weaker.  During the
ensuing ``flash mixing'', helium and carbon from the core will be carried
outward into the envelope while hydrogen from the envelope will be mixed
into the core.

\begin{figure}[t]
\centerline{\includegraphics[width=25pc]{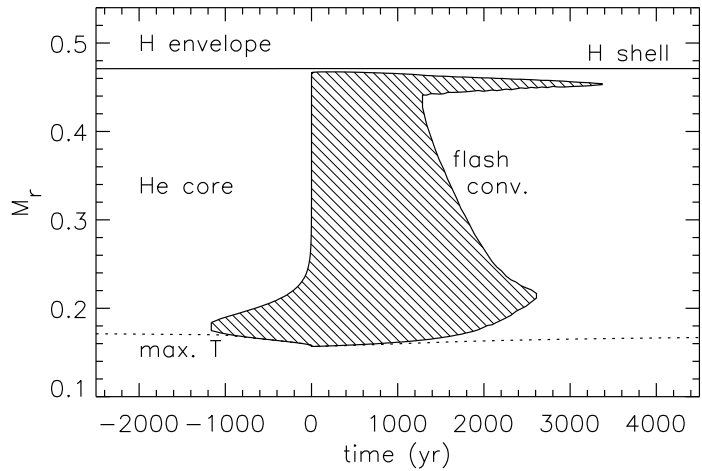}}
\caption{Time dependence of the flash convection zone (shaded area)
during a canonical helium flash.  The zero-point of the timescale
corresponds to the peak of the flash.  The ordinate gives
the mass coordinate $M_r$ in solar units.  The helium flash occurs
off-center at the point of maximum temperature (dotted curve).  During
a canonical flash the flash convection does not reach into the
hydrogen envelope, and the surface composition of the star is unchanged.}
\end{figure}

Our calculations have revealed two types of flash mixing, depending
on where the flash occurs along the white-dwarf cooling curve
(see \opencite{Lanz03} for a fuller discussion):

\begin{figure}[t]
\centerline{\includegraphics[width=25pc]{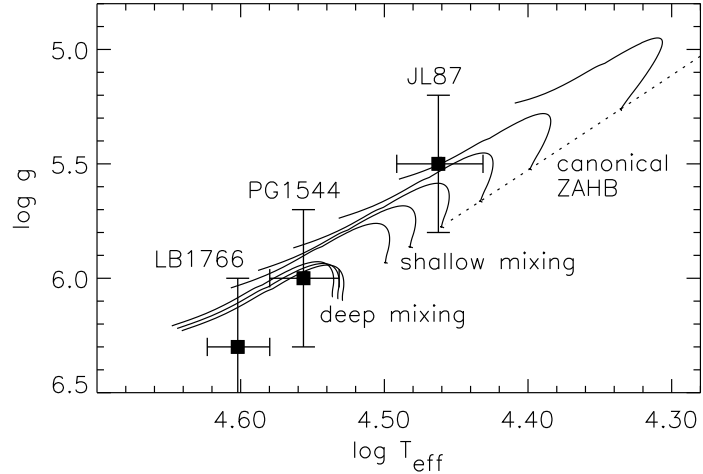}}
\caption{Comparison of the stellar parameters for the He-rich sdB
stars PG1544+488, JL87 and LB1766 with EHB evolutionary tracks for
canonical models (reddest 4 tracks), models with shallow
mixing (intermediate 2 tracks) and models with deep
mixing (bluest 3 tracks).  The dotted line denotes the canonical
ZAHB.}
\end{figure}

\begin{itemize}[$\bullet$]
\item ``Deep'' mixing:  If flash mixing occurs
when the core is fully convective outside the flash site,
then the envelope H will be mixed into the hot
He-burning regions and rapidly burned.  The resulting EHB star
will be enriched in He and C and greatly depleted in H.
\vspace{-0.03in}
\item ``Shallow'' mixing:  If flash mixing occurs later when there
is a distinct convective shell in the outer part of the core,
the envelope H will not be mixed deeply and thus
will not be burned.  The EHB star
will again be He- and C-rich but will also have significant H.
\end{itemize}

\vspace{-0.03in}
Figure 2 compares the stellar parameters for our three He-rich
sdB stars with EHB evolutionary tracks with
flash mixing.  Within the errors both the stellar
parameters and surface compositions of PG1544+488 and JL87 agree
with those predicted for stars with deep and shallow flash mixing,
respectively.   Flash mixing may therefore
represent a new evolutionary channel for producing He-rich sdB with
enhanced C.  However, LB1766 must have had a very
different evolutionary history.

\end{article}
\end{document}